\documentclass[prd,superscriptaddress,amsmath,notitlepage,twocolumn,nofootinbib]{revtex4-1}

\pdfoutput=1

\usepackage[utf8]{inputenc}
\usepackage{microtype}

\usepackage{adjustbox}
\usepackage{graphicx}
\usepackage{float}
\usepackage{epstopdf,cancel}
\usepackage{epsf,latexsym,bbm,euscript}
\usepackage{amsmath,amssymb}
\usepackage{mathtools}
\usepackage{times,graphics}
\usepackage{soul,xcolor}
\usepackage{mathtools}
\usepackage{tensor}
\usepackage{array}
\usepackage{tabularx}

\usepackage{booktabs}
\newcommand{\RNum}[1]{\uppercase\expandafter{\romannumeral #1\relax}}

\makeatletter
\g@addto@macro\bfseries{\boldmath}
\makeatother

\makeatletter
\newcommand*{\defeq}{\mathrel{\rlap{%
	\raisebox{0.3ex}{$\m@th\cdot$}}%
	\raisebox{-0.3ex}{$\m@th\cdot$}}%
	=}
\newcommand*{\eqdef}{=\mathrel{\rlap{%
	\raisebox{0.3ex}{$\m@th\cdot$}}%
	\raisebox{-0.3ex}{$\m@th\cdot$}}%
	}
\makeatother

\definecolor{ultramarine}{rgb}{0.07, 0.04, 0.56}

\def\pd{{\partial}}

\def\sg{\textsl{g}}

\def\eE{\EuScript{E}}
\def\eF{\EuScript{F}}
\def\eL{\EuScript{L}}

\newcommand{\sumj}{\sum\limits_{j \geqslant \frac{1}{2}}^\infty}
\newcommand{\overbar}[1]{\mkern 1.5mu\overline{\mkern-1.5mu#1\mkern-1.5mu}\mkern 1.5mu}

\usepackage{url,hyperref}
\hypersetup{colorlinks,linkcolor={blue!55!black},citecolor={red!45!black},urlcolor={blue!45!black},breaklinks=true}

\allowdisplaybreaks

\begin{document}

\title{Physical black holes in fourth-order gravity}

\author{Sebastian Murk}
\email{sebastian.murk@mq.edu.au}
\affiliation{Department of Physics and Astronomy, Macquarie University, Sydney, New South Wales 2109, Australia}
\affiliation{Sydney Quantum Academy, Sydney, New South Wales 2006, Australia}

\begin{abstract}
The existence of black holes is a central prediction of general relativity and thus serves as a basic consistency test for modified theories of gravity. In spherical symmetry, only two classes of dynamic solutions are compatible with the formation of an apparent horizon in finite time of a distant observer. Moreover, the formation of black holes follows a unique scenario involving both types of solutions. To be compatible with their existence, any self-consistent theory of modified gravity must satisfy several constraints. We derive properties of the modified gravity terms of $\mathfrak{f}(R)$ and generic fourth-order gravity theories and find that they naturally accommodate both classes of solutions. Consequently, the observation of an apparent horizon by itself may not suffice to distinguish between general relativity and modifications including up to fourth-order derivatives in the metric.
\end{abstract}

\maketitle

\section{Introduction} \label{sec:intro}
Black holes are arguably the most celebrated prediction of general relativity (GR). Due to spectacular advances in observational astronomy, strong evidence for the existence of dark massive compact objects (so-called astrophysical black holes) has accumulated over the last few decades, thus gradually shifting our perception of black holes from purely mathematical curiosities to real physical entities.

GR has so far managed to withstand all experimental tests. Nevertheless, its perceived shortcomings (e.g.\ the presence of singularities) have motivated the development and study of various modified theories of gravity (MTG), i.e.\ extensions and/or generalizations of GR involving additional gravitational degrees of freedom, typically through the inclusion of higher-order curvature corrections \cite{cdlno:09,blldc:18}\footnote{In the interest of brevity, many other appealing applications of MTG have been omitted from the discussion here. For a comprehensive overview, the interested reader may wish to consult Refs.~\cite{cdl:11,no:07,noo:17,dft:10,sf:10}.}. In addition, theoretical considerations indicate that GR represents the low-energy limit of some effective field theory of quantum gravity \cite{dh:15,b:04,hv:book}.

To be considered a viable candidate theory, any proposed modification of GR must be compatible with current astrophysical and cosmological data. In particular, it must provide a model to describe the observed astrophysical black hole candidates, which are described as ultra-compact objects with or without a horizon in popular contemporary models \cite{bcns:19}. While there is no unanimously agreed upon definition of a black hole, its most commonly accepted feature is the presence of a trapped region \cite{c:19}, i.e.\ a spacetime domain where both ingoing and outgoing future-directed null geodesics originating from a two-dimensional spacelike surface with spherical topology have negative expansion. Its evolving outer boundary is the apparent horizon. Following the nomenclature of Frolov \cite{f:14}, we refer to a trapped region bounded by an apparent horizon as a physical black hole (PBH). A PBH may include characteristic features of classical black hole solutions, such as an event horizon or singularity, or it may be a singularity-free regular black hole. Unlike the global notion of an event horizon, the apparent horizon is a well-defined quasilocal observable, i.e.\ its presence or absence is (at least in principle) detectable through quasilocal measurements \cite{v:14}. This makes it a suitable tool for practical purposes. In particular, it brings about the question of whether the apparent horizon can be used as a means to observationally distinguish between GR and various alternative theories of gravity \cite{cp:19}. To be of physical relevance, the apparent horizon must form in finite time according to a distant observer (Bob) \cite{mt:21c,dmt:21}. 

It is natural to ask whether the existence of PBHs as defined above, i.e.\ formation of an apparent horizon in finite time of Bob, imposes constraints on the mathematical structure of the modified Einstein equations in various MTG. A recent analysis \cite{mt:21b} has identified several such constraints for arbitrary metric MTG. Our goal is to determine whether families of higher-order gravity theories with up to fourth-order derivatives in the metric are compatible with the PBHs of semiclassical gravity, or if PBH solutions in these theories --- if they exist at all --- must have a fundamentally different mathematical structure. The fourth-order gravity theories we consider are particularly interesting in the context of quantum gravity since they are renormalizable \cite{s:77,sf:10}. Their field equations contain various additional degrees of freedom, including a massive scalar mode as well as massless and massive spin-2 vector modes.

We find that fourth-order MTG naturally include the PBH solutions of semiclassical gravity, i.e.\ no additional constraints are required to be compatible with the formation of an apparent horizon in finite time of Bob. This implies that the semiclassical solutions correspond to zeroth-order terms in perturbative solutions of these models, and the observation of an apparent horizon by itself is not a distinguishing feature between GR and various higher-order MTG models.

The remainder of this article is organized as follows: in Sec.~\ref{sec:prerequisites}, we discuss mathematical prerequisites and introduce key concepts of our analysis. In Sec.~\ref{sec:PBHsol}, we briefly summarize the properties of PBH solutions in semiclassical gravity. In Sec.~\ref{sec:MTG}, we describe the methodology of our effective field theory approach and review the constraints MTG must satisfy to be compatible with semiclassical PBHs. We then investigate the constraints in the context of $\mathfrak{f}(R)$ gravity (Sec.~\ref{sec:f(R)}) and generic fourth-order gravity theories (Sec.~\ref{sec:fog}) and derive properties of their respective modified gravity terms. Lastly, we discuss the implications of our findings and motivate avenues for further research (Sec.~\ref{sec:disc}).

\section{Prerequisites and general considerations} \label{sec:prerequisites}
Throughout this article, we use the $(-+++)$ signature of the metric $\tensor{\sg}{_\mu_\nu}$ and work in units where $\hbar = c = G = 1$. Working in the framework of semiclassical gravity, we use classical notions (e.g.\ metric, horizons, etc.) and describe dynamics via the semiclassical Einstein equations $\tensor{G}{_\mu_\nu} = 8 \pi \tensor{T}{_\mu_\nu}$ or modifications thereof, where $\tensor{T}{_\mu_\nu} \equiv \langle \tensor{\hat{T}}{_\mu_\nu} \rangle_\omega$ denotes the expectation value of the renormalized energy-momentum tensor (EMT) that describes the entire matter content, i.e.\ both the collapsing matter and the produced excitations of the quantum fields. We do not make any assumptions about the matter content of any given theory, the quantum state $\omega$, or the underlying reason(s) for modifications of the gravitational Lagrangian density $\eL_\text{g}$, which we organize according to powers of derivatives in the metric, i.e.\
\begin{align}
		\eL_\text{g} & = \frac{{M_\text{Pl}}^2}{16\pi} \big( R + \lambda \eF(\tensor{\sg}{^\mu^\nu}, \tensor{R}{_\mu_\nu_\rho_\sigma}) \big) \label{eq:gravLagr} \\
		& \hspace*{-5mm} = \frac{{M_\text{Pl}}^2}{16\pi} R + a_1 R^2 + a_2 \tensor{R}{_\mu_\nu} \tensor{R}{^\mu^\nu} + a_3 \tensor{R}{_\mu_\nu_\rho_\sigma} \tensor{R}{^\mu^\nu^\rho^\sigma} + \ldots , \nonumber
\end{align}
where the cosmological constant term was omitted, $M_\text{Pl}$ is the Planck mass that we set to one in what follows, and the coefficients $a_1$, $a_2$, $a_3$ are dimensionless. The dimensionless parameter $\lambda$ sets the scale of our perturbative analysis (see Sec.~\ref{sec:MTG}) and is set to one at the end of our calculations. 

In $(3+1)$ dimensions, the Einstein--Hilbert action 
\begin{align}
	\mathcal{S}_{\text{EH}} &= \frac{1}{16 \pi} \int \sqrt{-\sg} \; R \; d^4 x
	\label{eq:EHaction}
\end{align}
of classical GR is the most general gravitational action that can be constructed from symmetric rank 2 tensors involving at most second-order derivatives in the metric while maintaining diffeomorphism invariance (which implies, \textit{inter alia}, that the EMT is divergence-free, i.e.\ $\nabla_\mu \tensor{T}{^\mu^\nu} = 0$). Here, $\sg \equiv \det (\tensor{\sg}{_\mu_\nu})$ denotes the determinant of the metric tensor, and the gravitational Lagrangian density $\eL_\text{g}$ is strictly linear in the Ricci scalar $R$, but this is no longer true for the higher-derivative MTG we consider in Sec.~\ref{sec:f(R)} and Sec.~\ref{sec:fog}.

We restrict our considerations to spherical symmetry. In Schwarzschild coordinates, a general spherically symmetric metric is given by
\begin{align}
	ds^2 = - e^{2h(t,r)} f(t,r) dt^2 + f(t,r)^{-1} dr^2 + r^2 d\Omega ,
	\label{eq:metric}
\end{align}
where $r$ denotes the areal radius, the Misner--Sharp mass \cite{ms:64} $C(t,r)/2$ is invariantly defined via 
\begin{align}
	f(t,r) \defeq \partial_\mu r \partial^\mu r = 1 - C(t,r)/r ,
	\label{eq:f}
\end{align}
and the function $h(t,r)$ plays the role of an integrating factor in coordinate transformations, e.g.\ 
\begin{align}
	dt = e^{-h} \left( e^{h_+} dv - f^{-1} dr \right) 
\end{align}
between Schwarzschild $(t,r)$ and advanced null $(v,r)$ coordinates. The definition of Eq.~\eqref{eq:f} is particularly convenient as it allows for a consistent description of solutions in four- and higher-dimensional models of both GR and MTG. The apparent horizon is located at the Schwarzschild radius $r_\sg(t)$, which corresponds to the largest root of $f(t,r)=0$ \cite{fefhm:17}. Its definition implies
\begin{align}
	C(t,r) &= r_\sg(t) + W(t,x) ,
	\label{eq:C}
\end{align}
where $x \defeq r - r_\sg$ denotes the coordinate distance from the apparent horizon, and
\begin{align}
	W(t,0) = 0 , \qquad W(t,x) < x \quad \forall \; x>0 .
\end{align}
In general, this definition is an observer-dependent notion. However, in spherical symmetry the apparent horizon is unambiguously defined in all foliations that respect this symmetry \cite{fefhm:17}. 

Apart from spherical symmetry, our only assumption is that a regular apparent horizon forms in finite time of a distant observer (Bob). Regularity is a necessary requirement to maintain predictability of the theory \cite{hv:book,pt:book}, and finite-time formation according to Bob is needed to ensure that the PBH solutions we consider are observationally relevant physical objects (as opposed to mere mathematical idealizations) \cite{mt:21c,dmt:21}. Mathematically, regularity of the horizon is expressed through the finiteness of the curvature scalars
\begin{align}
	\mathrm{T} & \defeq \tensor{T}{^\mu_\mu} = - R / 8\pi + \mathcal{O}(\lambda) , 
	\label{eq:cs_trace} \\
	\mathfrak{T} & \defeq \tensor{T}{^\mu^\nu}\tensor{T}{_\mu_\nu} = \tensor{R}{^\mu^\nu}\tensor{R}{_\mu_\nu} / 64 \pi^2 + \mathcal{O}(\lambda^2) ,
	\label{eq:cs_square}
\end{align}
i.e.\ the trace and square of the EMT, at the horizon. Our self-consistent approach is based on the assumption of at least continuity of the curvature invariants, but uses Schwarzschild coordinates where the metric is discontinuous. Imposing the requirement of regularity then allows us to identify the valid PBH solutions \cite{bmmt:19,mt:21a}.

It is convenient to work with the effective EMT components
\begin{align}
	\tensor{\tau}{_t} \defeq e^{-2h} \tensor{T}{_t_t} , \quad \tensor{\tau}{_t^r} \defeq e^{-h} \tensor{T}{_t^r} , \quad \tensor{\tau}{^r} \defeq \tensor{T}{^r^r} .
	\label{eq:effEMT}
\end{align}
The regularity requirement can then be expressed as
\begin{align}
	\mathrm{T} &= \left( \tensor{\tau}{^r} - \tensor{\tau}{_t} \right) / f \; \to \; g_1(t) f^{k_1} , 
	\label{eq:reg1} \\
	\mathfrak{T} &= \left[ \left( \tensor{\tau}{_t} \right)^2 - 2 \left( \tensor{\tau}{_t^r} \right)^2 + \left( \tensor{\tau}{^r} \right)^2 \right] / f^2 \; \to \; g_2(t) f^{k_2} ,
	\label{eq:reg2}
\end{align}
for some functions $g_{1,2}(t)$ and $k_{1,2} \geqslant 0$. \textit{A priori}, there are infinitely many solutions that satisfy these constraints. However, it has been demonstrated that only two distinct classes of solutions with $k_1 = k_2 \eqdef k$, $k \in \lbrace 0,1 \rbrace$, are admissible \cite{t:20,mt:21a}. We briefly summarize their properties in Sec.~\ref{sec:PBHsol}.

\section{Physical black holes in semiclassical gravity} \label{sec:PBHsol}
In spherical symmetry, the semiclassical Einstein equations for the components $\tensor{G}{_t_t}$, $\tensor{G}{_t^r}$, and $\tensor{G}{^r^r}$ are given by
\begin{align}
	\partial_r C &= 8 \pi r^2 \tensor{\tau}{_t} / f ,
	\label{eq:EEGRtt} \\
	\partial_t C &= 8 \pi r^2 e^{h} \tensor{\tau}{_t^r} ,
	\label{eq:EEGRtr} \\
	\partial_r h &= 4 \pi r \left( \tensor{\tau}{_t} + \tensor{\tau}{^r} \right) / f^2 .
	\label{eq:EEGRrr}
\end{align}

\onecolumngrid
\vspace*{4mm}
\begin{table}[!htpb] 
	\centering
	\resizebox{0.99\textwidth}{!}{ 
		\begin{tabular}{ >{\raggedright\arraybackslash}m{0.15\linewidth} | @{\hskip 0.025\linewidth} >{\raggedright\arraybackslash}m{0.385\linewidth} @{\hskip 0.05\linewidth} | @{\hskip 0.025\linewidth} >{\raggedright\arraybackslash}m{0.35\linewidth}} 
			& $k=0$ solutions & $k=1$ solution 
			\\ \toprule
			Metric functions 
			\vspace*{13.5mm}
			& 
			{\begin{flalign}
				C &= r_\sg - c_{12} \sqrt{x} + \sum\limits_{j \geqslant 1}^\infty c_j x^j & \label{teq:k0-C} \tag{k0.1} \\
				h &= - \frac{1}{2} \ln \frac{x}{\xi} + \sumj h_j x^j & \label{teq:k0-h} \tag{k0.2} 
			\end{flalign}} 
			\quad & \quad 
			{\begin{flalign}
				C &= r_\sg + x - c_{32} x^{3/2} + \sum\limits_{j \geqslant 2}^\infty c_j x^j & \label{teq:k1-C} \tag{k1.1}  \\
				h &= - \frac{3}{2} \ln \frac{x}{\xi} + \sumj h_j x^j & \label{teq:k1-h} \tag{k1.2} 
			\end{flalign}} 
			\\[-4mm]
			Leading coefficient 
			\vspace*{5.5mm}
			& 
			{\begin{flalign}
				c_{12}&= 4 \sqrt{\pi} r_\sg^{3/2} \Upsilon & \label{teq:k0-c12} \tag{k0.3}
			\end{flalign}} 
			\quad & \quad 
			{\begin{flalign}
				c_{32}&= 4 r_\sg^{3/2} \sqrt{- \pi e_2 / 3} & \label{teq:k1-c32} \tag{k1.3}
			\end{flalign}} 
			\\[-4mm]  		
			Horizon dynamics 
			\vspace*{5.5mm}
			& 
			{\begin{flalign}
				r_\sg^\prime &= \pm c_{12} \sqrt{\xi} / r_\sg & \label{teq:k0-rg} \tag{k0.4}
			\end{flalign}} 
			\quad & \quad 
			{\begin{flalign}
				r_\sg^\prime &= \pm c_{32} \xi^{3/2} / r_\sg & \label{teq:k1-rg} \tag{k1.4}
			\end{flalign}} 
			\\[-4mm] 
			Effective EMT 
			\vspace*{20mm}
			& 
			{\begin{flalign}
				\tensor{\tau}{_t} &= - \Upsilon^2 + \sum\limits_{j \geqslant \frac{1}{2}}^\infty e_j x^j & \label{teq:k0-tautt} \tag{k0.5} \\
				\tensor{\tau}{_t^r} &= \pm \Upsilon^2 + \sum\limits_{j \geqslant \frac{1}{2}}^\infty \phi_j x^j & \label{teq:k0-tautr} \tag{k0.6} \\
				\tensor{\tau}{^r} &= - \Upsilon^2 + \sum\limits_{j \geqslant \frac{1}{2}}^\infty p_j x^j & \label{teq:k0-taurr} \tag{k0.7}
			\end{flalign}} 
			\quad & \quad 
			{\begin{flalign}
				\tensor{\tau}{_t} &= E f + \sum\limits_{j \geqslant 2}^\infty e_j x^i & \label{teq:k1-tautt} \tag{k0.5} \\
				\tensor{\tau}{_t^r} &= \Phi f + \sum\limits_{j \geqslant 2}^\infty \phi_j x^j & \label{teq:k1-tautr} \tag{k0.6} \\
				\tensor{\tau}{^r} &= P f + \sum\limits_{j \geqslant 2}^\infty p_j x^i & \label{teq:k1-taurr} \tag{k0.7} 
			\end{flalign}} 
			\\[-4mm] 
			Ricci scalar
			\vspace*{8mm}
			& 
			{\begin{flalign}
				R &= R_0 + R_{12} \sqrt{x} + R_1 x + \sum\limits_{j \geqslant \frac{3}{2}}^\infty R_j x^j & \label{teq:k0-R} \tag{k0.8}
			\end{flalign}} 
			\quad & \quad 
			{\begin{flalign}
				R &= 2/r_\sg^2 + R_1 x + \sum\limits_{j \geqslant \frac{3}{2}}^\infty R_j x^j & \label{teq:k1-R} \tag{k1.8}
			\end{flalign}} 
			\\
			\bottomrule
		\end{tabular}
	}
	\caption{Comparison of the two classes of dynamic solutions in spherical symmetry. The metric functions $C$ and $h$ [cf.\ Eqs.~\eqref{eq:metric}--\eqref{eq:f}] are obtained as the solutions of Eqs.~\eqref{eq:EEGRtt} and \eqref{eq:EEGRrr} and are written together with the effective EMT components and Ricci scalar as series expansions in terms of the coordinate distance $x \defeq r - r_\sg$ from the apparent horizon $r_\sg$. The function $\Upsilon(t)>0$ parametrizes the leading contributions to the effective EMT components for $k=0$ solutions, and $\xi(t)$ is determined by the choice of time variable. In spherical symmetry, the geometry near the apparent horizon \cite{bmmt:19,t:20,mt:21a} is constrained sufficiently enough to identify $\Upsilon(t)$ and $\xi(t)$ and match them with the semiclassical results \cite{mmt:21a}. The letter $j \in \mathbb{Z} \frac{1}{2}$ labels half-integer and integer coefficients and powers of $x$. Since only the leading terms in each series are relevant, we simplify the notation by writing $c_{12}$ instead of $c_{1/2}$, and similarly for higher orders and coefficients of the EMT expansion and Ricci scalar. To remind us of their connection to physical quantities, the coefficients of the effective EMT components are denoted $e_j$ (energy density), $\phi_j$ (flux), and $p_j$ (pressure). Consistency of Eqs.~\eqref{eq:EEGRtr}--\eqref{eq:EEGRrr} implies $E = - P = 1/(8 \pi r_\sg^2)$ and $\Phi=0$. The lower (upper) signature in Eqs.~\eqref{teq:k0-rg}, \eqref{teq:k0-tautr}, and \eqref{teq:k1-rg} describes an evaporating PBH (an expanding white hole). The dynamic behavior of the horizon $r_\sg^\prime \defeq dr_\sg/dt$ is determined by Eq.~\eqref{eq:EEGRtr}, and also implicitly through the requirement that the Ricci scalar $R$ be finite at the horizon, that is Eqs.~\eqref{teq:k0-rg} and \eqref{teq:k1-rg} must hold for the $k=0$ and $k=1$ solutions, respectively, in order for the divergent terms in the series expansion of $R$ to vanish. The Einstein equations Eqs.~\eqref{eq:EEGRtt}--\eqref{eq:EEGRrr} hold order by order in terms of $x$. Accordingly, explicit expressions for higher-order terms in the metric functions are obtained by matching those of the same order in the EMT expansion \cite{bmt:19,mt:21b}.} 
	\label{tab:PBHsol} 
\end{table}
\vspace*{4mm}
\twocolumngrid

Only two distinct classes of dynamic ($r_\sg^\prime(t) \defeq d r_\sg / dt \neq 0$) solutions are compatible with the formation of an apparent horizon in finite time of a distant observer (Bob). With respect to the regularity conditions of Eqs.~\eqref{eq:reg1}--\eqref{eq:reg2}, they correspond to the values $k=0$ (i.e.\ $k_1 = k_2 = 0$) and $k = 1$ (i.e.\ $k_1 = k_2 = 1$). Their main properties are summarized in Tab.~\ref{tab:PBHsol}. For a detailed derivation, the reader is referred to Refs.~\cite{mt:21a,t:20}. For the class of $k=0$ solutions, a static solution is impossible, as in this case $\mathfrak{T}$ would diverge at the apparent horizon. For $k=1$ on the other hand, a static solution is possible, but there is only one self-consistent dynamic solution [described by Eqs.~\eqref{teq:k1-C}--\eqref{teq:k1-R}] for which the energy density $E \defeq - \tensor{T}{^t_t}$ and pressure $P \defeq \tensor{T}{^r_r}$ at the horizon take on their extreme values $E = - P = 1/(8 \pi r_\sg^2)$. For this extreme-valued $k=1$ solution, the identity $c_2 = c_{32} h_{12}$ between the coefficients of its metric functions Eqs.~\eqref{teq:k1-C}--\eqref{teq:k1-h} holds (a derivation is provided in Sec.~IV.B.\ of Ref.~\cite{mt:21b}) and leads to many simplifying cancellations, e.g.\ the vanishing of the $\sqrt{x}$ term in the Ricci scalar expansion [see Eq.~\eqref{teq:k1-R}] due to the fact that $R_{12} \propto c_{32} h_{12} - c_2$.

When the matter content of a theory is not specified explicitly, the permissible states of matter are usually constrained by means of energy conditions \cite{he:book,mm:book,ks:20}. The weakest of all energy conditions is the null energy condition (NEC), which postulates that $\tensor{T}{_\mu_\nu} \tensor{\ell}{^\mu} \tensor{\ell}{^\nu} \geqslant 0$, i.e.\ the contraction of the EMT with any null vector $\tensor{\ell}{^\mu}$ is non-negative. In contrast to classical spherically symmetric solutions, the NEC is violated in the vicinity of the outer apparent horizon for both classes (i.e.\ $k=0$ and $k=1$) of semiclassical solutions \cite{dmt:21,bhl:18,bmmt:19,t:20}. For expanding white hole solutions ($r_\sg^\prime(t) > 0$), the energy density, pressure, and flux perceived by an infalling observer diverge fast enough to violate quantum energy inequalities that bound violations of the NEC \cite{fp:06,ko:15,f:book}, indicating either the breakdown of semiclassical physics or confirming the instability of white hole horizons \cite{mmt:21b}. As our interest lies in describing scenarios resulting from gravitational collapse, we restrict our considerations to evaporating PBH solutions ($r_\sg^\prime(t) < 0$) in what follows.

At the instant of its formation, a PBH is described by the extreme-valued $k=1$ solution \cite{mt:21a}. Its behavior then immediately switches to that of a $k=0$ solution. Since the energy density and pressure are negative in the vicinity of the outer apparent horizon and positive in the vicinity of the inner horizon \cite{t:19,f:14}, these two quantities jump at the intersection of the two horizons. However, the abrupt transition from $f^1$ to $f^0$ behavior is only of conceptual importance as this aspect of the evolution is continuous in $(v,r)$ coordinates, and there is no discontinuity according to observers crossing the respective horizon surfaces.

\section{Modified gravity: Einstein equations and constraints} \label{sec:MTG}
Variation of the gravitational action results in
\begin{align}
	\tensor{G}{_\mu_\nu} + \lambda \tensor{\EuScript{E}}{_\mu_\nu} = 8 \pi \tensor{T}{_\mu_\nu} ,
	\label{eq:mEE}
\end{align}
where $\tensor{G}{_\mu_\nu}$ denotes the Einstein tensor, and the terms $\tensor{\EuScript{E}}{_\mu_\nu}$ result from the variation of $\eF(\tensor{\sg}{^\mu^\nu}, \tensor{R}{_\mu_\nu_\rho_\sigma})$ [cf.\ Eq.~\eqref{eq:gravLagr}]. In spherical symmetry, the modified Einstein equations take the form
\begin{align}
	& f r^{-2} e^{2h} \partial_r C + \lambda \tensor{\EuScript{E}}{_t_t} = 8 \pi \tensor{T}{_t_t} ,
	\label{eq:mEEtt} \\
	& r^{-2} \partial_t C + \lambda \tensor{\EuScript{E}}{_t^r} = 8 \pi \tensor{T}{_t^r} ,
	\label{eq:mEEtr} \\
	& 2 f^2 r^{-1} \partial_r h - f r^{-2} \partial_r C + \lambda \tensor{\EuScript{E}}{^r^r} = 8 \pi \tensor{T}{^r^r} .
	\label{eq:mEErr}
\end{align}
We assume that there is a solution of Eq.~\eqref{eq:mEE} with the metric functions 
\begin{align}
	C_\lambda & \eqdef \bar{C}(t,r) + \lambda \Sigma(t,r) , 
	\label{eq:C_lambda} \\
	h_\lambda & \eqdef \bar{h}(t,r) + \lambda \Omega(t,r) ,
	\label{eq:h_lambda}
\end{align}
where the bar labels functions of semiclassical gravity described in Sec.~\ref{sec:PBHsol}, e.g.\ $\bar{C} \defeq r_\sg + \bar{W}$ [cf.\ Eq.~\eqref{eq:C}], and $\Sigma$ and $\Omega$ denote the perturbative corrections. To avoid artifactual divergences, we use the physical value of $r_\sg(t)$ that corresponds to the perturbed metric $\sg_\lambda \defeq \tensor{\bar{\sg}}{_\mu_\nu} + \lambda \tensor{\tilde{\sg}}{_\mu_\nu}$, i.e.\ $C_\lambda(t,r_\sg) = r_\sg$. Similarly, the EMT depends on $\lambda$ through the metric $\sg_\lambda$, and potentially also through effective corrections resulting from perturbative corrections to the modified field equations Eqs.~\eqref{eq:mEEtt}--\eqref{eq:mEErr}. It is decomposed as
\begin{align}
	\tensor{T}{_\mu_\nu} \eqdef \tensor{\bar{T}}{_\mu_\nu} + \lambda \tensor{\tilde{T}}{_\mu_\nu} ,
	\label{eq:MTG_EMT}
\end{align}
where $\tensor{\bar{T}}{_\mu_\nu} \equiv \tensor{T}{_\mu_\nu}[\bar{C},\bar{h}]$ corresponds to the semiclassical term. The perturbative corrections must satisfy the boundary conditions
\begin{align}
	& \Sigma(t,0) = 0 , \\
	& \hspace*{-1.2mm} \lim_{r \to r_\sg} \Omega(t,r)/\bar{h}(t,r) = \mathcal{O}(1) ,
\end{align}
where the first condition follows from the definition of the horizon radius $r_\sg$, and the second condition ensures that perturbations can be treated as small, i.e.\ the divergence of $\Omega$ on approach to the horizon (i.e.\ as $r \to r_\sg$) must not be stronger than that of $\bar{h}$.
Substituting Eqs.~\eqref{eq:C_lambda}--\eqref{eq:h_lambda} into Eq.~\eqref{eq:mEE} and keeping only first-order terms in $\lambda$ results in
\begin{align}
	\tensor{\bar{G}}{_\mu_\nu} + \lambda \tensor{\tilde{G}}{_\mu_\nu} + \lambda \tensor{\bar{\eE}}{_\mu_\nu} = 8 \pi \left( \tensor{\bar{T}}{_\mu_\nu} + \lambda \tensor{\tilde{T}}{_\mu_\nu} \right) ,
\end{align}
where $\tensor{\bar{G}}{_\mu_\nu} \equiv \tensor{G}{_\mu_\nu}[\bar{C},\bar{h}]$, $\tensor{\tilde{G}}{_\mu_\nu}$ corresponds to the first-order term of the Taylor expansion in $\lambda$ where each monomial involves either $\Sigma$ or $\Omega$, and $\tensor{\bar{\eE}}{_\mu_\nu} \equiv \tensor{\eE}{_\mu_\nu}[\bar{C},\bar{h}]$, i.e.\ the modified gravity terms are functions of the unperturbed solutions.

If we adopt the schematic separation of the EMT according to Eq.~\eqref{eq:MTG_EMT} for the effective EMT components defined in Eq.~\eqref{eq:effEMT}, i.e.\ $\tau = \bar{\tau} + \lambda \tilde{\tau}$, then the modified gravity field equations Eqs.~\eqref{eq:mEEtt}--\eqref{eq:mEErr} can be written explicitly as \cite{mt:21b}
\begin{alignat}{2}
	- & \Sigma \partial_r \bar{C} + \left( r - \bar{C} \right) \partial_r \Sigma + r^3 e^{- 2 \bar{h}} \tensor{\bar{\EuScript{E}}}{_t_t} = 8 \pi r^3 \tensor{\tilde{\tau}}{_t} , 
	\label{eq:mGravEFEtt} \\
	& \partial_t \Sigma + r^2 \tensor{\bar{\EuScript{E}}}{_t^r} = 8 \pi r^2 e^{\bar{h}} ( \Omega \tensor{\bar{\tau}}{_t^r} + \tensor{\tilde{\tau}}{_t^r}) , 
	\label{eq:mGravEFEtr} \\
	\begin{split}
		& \Sigma \partial_r \bar{C} -  ( r - \bar{C}) (4\Sigma \partial_r \bar{h} + \partial_r \Sigma) \\
		& \qquad \hspace{1.65mm} + 2( r - \bar{C})^2 \partial_r \Omega + r^3 \tensor{\bar{\EuScript{E}}}{^r^r} = 8 \pi r^3 \tensor{\tilde{\tau}}{^r} . 
		\label{eq:mGravEFErr}
	\end{split}
\end{alignat}
To be compatible with the dynamic PBH solutions of semiclassical gravity (summarized in Sec.~\ref{sec:PBHsol}, Tab.~\ref{tab:PBHsol}), any arbitrary MTG must satisfy several constraints. They are summarized in Tab.~\ref{tab:MTGcc} and derived explicitly in Ref.~\cite{mt:21b}. First, the series expansions of the MTG terms $\tensor{\bar{\eE}}{_\mu_\nu}$ in terms of $x \defeq r - r_\sg$ must conform to the structures prescribed by Eqs.~\eqref{teq:k0-Ett}--\eqref{teq:k0-Err} and Eqs.~\eqref{teq:k1-Ett}--\eqref{teq:k1-Err} for the $k=0$ and $k=1$ solutions, respectively. Second, the coefficients of the MTG terms must satisfy three additional identities [Eqs.~\eqref{teq:k0-cc1cc2}--\eqref{teq:k0-cc3}] in the $k=0$ case and two additional relations [Eqs.~\eqref{teq:k1-cc1}--\eqref{teq:k1-cc2}] for the unique $k=1$ solution.

There is \textit{a priori} no reason to believe that the constraints imposed by Eqs.~\eqref{teq:k0-Ett}--\eqref{teq:k0-cc3} and Eqs.~\eqref{teq:k1-Ett}--\eqref{teq:k1-cc2} should or should not be satisfied in any particular MTG. If they are not satisfied, the MTG in question may still possess solutions corresponding to PBHs, albeit their mathematical structure must then be fundamentally different from those of semiclassical gravity (which may or may not give rise to observationally distinguishable features). On the other hand, if the constraints are satisfied identically in a particular MTG, then the semiclassical PBH solutions can be regarded as zeroth-order terms in perturbative solutions of this model. It is also possible that the constraints are satisfied only if additional conditions are fulfilled. In this case, compatibility with semiclassical PBHs would impose further constraints on the MTG in question, i.e.\ beyond those listed in Tab.~\ref{tab:MTGcc}. It is worth noting that, since semiclassical PBHs are described by the extreme-valued $k=1$ solution at their formation \cite{mt:21a}, the failure of a particular MTG to satisfy any one of the $k=1$ constraints Eqs.~\eqref{teq:k1-Ett}--\eqref{teq:k1-cc2} suffices to necessitate that the PBH formation scenario in that theory differs from that of semiclassical gravity.

In Sec.~\ref{sec:f(R)} and Sec.~\ref{sec:fog}, we examine the constraints in the context of $\mathfrak{f}(R)$ and generic fourth-order theories of gravity to determine which of the outcomes described above is realized and derive properties of their respective MTG terms.

\onecolumngrid
\vspace*{4mm}
\begin{table}[!htpb] 
	\centering
	\resizebox{\textwidth}{!}{ 
		\begin{tabular}{ >{\raggedright\arraybackslash}m{0.12\linewidth} | @{\hskip 0.025\linewidth} >{\raggedright\arraybackslash}m{0.355\linewidth} @{\hskip 0.03\linewidth} | @{\hskip 0.025\linewidth} >{\raggedright\arraybackslash}m{0.445\linewidth}} 
			& $k=0$ solutions & $k=1$ solution 
			\\ \toprule
			Decomposition of MTG terms 
			\vspace*{18.25mm}
			&
			{\begin{flalign}
				\tensor{\bar{\eE}}{_t_t} & = \frac{\ae_{\bar{1}}}{x} + \frac{\ae_{\overbar{12}}}{\sqrt{x}} + \ae_0 + \sumj \ae_j x^j &
				\label{teq:k0-Ett} \tag{k0.\RNum{1}} \\
				\tensor{\bar{\eE}}{_t^r} & = \frac{\oe_{\overbar{12}}}{\sqrt{x}} + \oe_0 + \sumj \oe_j x^j &
				\label{teq:k0-Etr} \tag{k0.\RNum{2}} \\
				\tensor{\bar{\eE}}{^r^r} & = \o_0 + \sumj \o_j x^j &
				\label{teq:k0-Err} \tag{k0.\RNum{3}}
			\end{flalign}} 
			\quad & \quad 
			{\begin{flalign}
				\tensor{\bar{\eE}}{_t_t} & = \frac{\ae_{\overbar{32}}}{x^{3/2}} + \frac{\ae_{\bar{1}}}{x} + \frac{\ae_{\overbar{12}}}{\sqrt{x}}  + \ae_0 + \sumj \ae_j x^j &
				\label{teq:k1-Ett} \tag{k1.\RNum{1}} \\
				\tensor{\bar{\eE}}{_t^r} & = \oe_0 + \sumj \oe_j x^j & 
				\label{teq:k1-Etr} \tag{k1.\RNum{2}} \\
				\tensor{\bar{\eE}}{^r^r} & = \sum\limits_{j \geqslant \frac{3}{2}}^\infty \o_j x^j &
				\label{teq:k1-Err} \tag{k1.\RNum{3}}
			\end{flalign}} 
			\\[-3mm] 
			Relations between MTG coefficients
			\vspace*{5.25mm}
			&
			{\begin{flalign}
				\ae_{\bar{1}} &= \sqrt{\bar{\xi}} \oe_{\overbar{12}} = \bar{\xi} \o_0 &
				\label{teq:k0-cc1cc2} \tag{k0.\RNum{4}} \\
				\ae_{\overbar{12}} &= 2 \sqrt{\bar{\xi}} \oe_0 - \bar{\xi} \o_{12} &
				\label{teq:k0-cc3} \tag{k0.\RNum{5}}
			\end{flalign}} 
			\quad & \quad 
			{\begin{flalign}
				\ae_{\overbar{32}} &= 2 \bar{\xi}^{3/2} \oe_0 - \bar{\xi}^3 \o_{32} \vphantom{\sqrt{\bar{\xi}}} &
				\label{teq:k1-cc1} \tag{k1.\RNum{4}} \\
				\ae_{\bar{1}} &= 2 \bar{\xi}^{3/2} \left( h_{12} \oe_0 + \oe_{12} \right) - \bar{\xi}^3 \left( 2 h_{12} \o_{32} + \o_2 \right) \vphantom{\sqrt{\bar{\xi}}} &
				\label{teq:k1-cc2} \tag{k1.\RNum{5}}
			\end{flalign}} 
			\\
			\bottomrule
		\end{tabular}
	} 
	\caption{Necessary conditions for the existence of semiclassical PBHs in arbitrary metric MTG. To be compatible with semiclassical PBHs of the $k=0$ ($k=1$) type, the MTG terms of arbitrary metric MTG must conform to the structures prescribed by Eqs.~\eqref{teq:k0-Ett}--\eqref{teq:k0-Err} [Eqs.~\eqref{teq:k1-Ett}--\eqref{teq:k1-Err}] when expanded in terms of $x \defeq r - r_\sg$. Additionally, their lowest-order coefficients must satisfy the three (two) identities given by Eqs.~\eqref{teq:k0-cc1cc2}--\eqref{teq:k0-cc3} [Eqs.~\eqref{teq:k1-cc1}--\eqref{teq:k1-cc2}].}
	\label{tab:MTGcc} 
\end{table}
\vspace*{4mm}
\twocolumngrid

\section{Physical black holes in $\mathfrak{f}(R)$ gravity} \label{sec:f(R)}
As mentioned in Sec.~\ref{sec:prerequisites}, the gravitational Lagrangian density of GR is strictly linear in the Ricci scalar $R$, i.e.\ $\eL_\text{g} = R$ [cf.\ Eq.~\eqref{eq:EHaction}]. One of the simplest conceivable modifications of GR is $\mathfrak{f}(R)$ gravity \cite{dft:10,sf:10}: a class of theories in which the linearity requirement is relaxed and $\eL_\text{g}$ is taken to be an arbitrary function of $R$, i.e.\ $\eL_\text{g} = \mathfrak{f}(R)$. Despite this seemingly simple modification, the relevant field equations in $\mathfrak{f}(R)$ theories already involve up to fourth-order derivatives in the metric. We consider generic $\mathfrak{f}(R)$ theories, i.e.\ $\mathfrak{f}(R) \eqdef R + \lambda \eF(R)$, where $\eF(R) = \beta R^q$ and $\beta, q \in \mathbb{R}$. For the action
\begin{align}
	\mathcal{S}_{\mathfrak{f}(R)} = \frac{1}{16 \pi} \int \big( \mathfrak{f}(R) + \eL_\text{m} \big) \sqrt{-\sg} \; d^4x + S_\text{b},
\end{align}
where the matter Lagrangian is represented by $\eL_\text{m}$ and $S_\text{b}$ denotes the boundary term, the field equations are given by
\begin{align}
	\mathfrak{f}' \tensor{R}{_\mu_\nu} - \frac{1}{2} \mathfrak{f} \tensor{\sg}{_\mu_\nu} + \left( \tensor{\sg}{_\mu_\nu} \square - \nabla_\mu \nabla_\nu \right) \mathfrak{f}' = 8 \pi \tensor{T}{_\mu_\nu} ,
\end{align}
where $\mathfrak{f}' \defeq \partial \mathfrak{f}(R) / \partial R$ and $\square \defeq \tensor{\sg}{^\mu^\nu} \nabla_\mu \nabla_\nu$ denotes the d'Alembertian. In spherical symmetry [i.e.\ for the metric of Eq.~\eqref{eq:metric}], it is given by
\begin{align}
	\square \eF^\prime &= \left[ \partial_t \partial^t + \partial_r \partial^r + \left( \partial_t  h \right) \partial^t +
 \left( \partial_r h + 2/r \right) \partial^r \right] \eF^\prime 
	. \label{eq:d'Alembertian}
\end{align}
Second-order covariant derivatives of a scalar function can be expressed in terms of partial derivatives, i.e.\
\begin{align}	
	\nabla_\mu \nabla_\nu \eF^\prime = \left( \partial_\mu \partial_\nu \ - \tensor{\Gamma}{^\zeta_\mu_\nu} \partial_\zeta \right) \eF^\prime
	. \label{eq:sec_cov_der}
\end{align}
The modified Einstein equations are then given by [cf.\ Eq.~\eqref{eq:mEE}]
\begin{align}
	\resizebox{.99\linewidth}{!}{$
	\tensor{G}{_\mu_\nu} + \lambda \Big[ \eF^\prime \tensor{R}{_\mu_\nu}
 - \frac{1}{2} \eF \tensor{\sg}{_\mu_\nu}+ \left( \tensor{\sg}{_\mu_\nu} \square - \nabla_\mu \nabla_\nu \right) \eF^\prime \Big]
  = 8 \pi \tensor{T}{_\mu_\nu} 
  . $} \label{eq:mEEe}
\end{align}
We obtain expressions for the modified gravity terms $\tensor{\bar{\EuScript{E}}}{_\mu_\nu}$ by performing the expansion in $\lambda$ and keeping terms only up to the first order, i.e.\ 
\begin{align}
	\tensor{\bar{\EuScript{E}}}{_\mu_\nu} = \eF^\prime \tensor{\bar{R}}{_\mu_\nu} - \frac{1}{2} \eF\, \tensor{\bar{\sg}}{_\mu_\nu} + \left( \tensor{\bar{\sg}}{_\mu_\nu} \bar{\square} - \bar{\nabla}_\mu \bar{\nabla}_\nu \right) \eF^\prime 
	, \label{eq:E_mu_nu}
\end{align}
where all objects labeled by the bar are evaluated with respect to the unperturbed metric $\tensor{\bar{\sg}}{_\mu_\nu}$, and $\eF \equiv \eF(\bar{R})$.

Using Eqs.~\eqref{eq:d'Alembertian}--\eqref{eq:sec_cov_der} to evaluate Eq.~\eqref{eq:E_mu_nu} with the metric of Eq.~\eqref{eq:metric}, we obtain the explicit form of the MTG terms $\tensor{\bar{\EuScript{E}}}{_\mu_\nu}$ as a function of unperturbed quantities, i.e.\
\begin{widetext}
\begin{align}
	\frac{\tensor{\bar{\eE}}{_t_t}}{\beta \lambda} &= 
	\eF^\prime \tensor{\bar{R}}{_t_t} - \frac{1}{2} \eF \tensor{\bar{\sg}}{_t_t} + \left[ \tensor{\bar{\sg}}{_t_t} \left( \pd_t \pd^t + \pd_r \pd^r + \left( \pd_t \bar{h} \right) \pd^t + \left( \pd_r \bar{h} + \frac{2}{r} \right) \pd^r \right) - \pd_t \pd_t + \tensor{\Gamma}{^t_t_t}  \pd_t + \tensor{\Gamma}{^r_t_t} \pd_r \right] \eF^\prime
	, \label{eq:EttF} \\
	\frac{\tensor{\bar{\eE}}{_t^r}}{\beta \lambda} &=
	\eF^\prime \tensor{\bar{R}}{_t^r} - \left( \pd_t \pd^r + \tensor{\Gamma}{^r_t_t} \pd^t + \tensor{\Gamma}{^r_t_r} \pd^r \right) \eF^\prime
	, \label{eq:EtrF} \\
	\frac{\tensor{\bar{\eE}}{^r^r}}{\beta \lambda} &=
	\eF^\prime \tensor{\bar{R}}{^r^r} - \frac{1}{2} \eF \tensor{\bar{\sg}}{^r^r} + \tensor{\bar{\sg}}{^r^r} \left[ \pd_t \pd^t + \left( \pd_t \bar{h} - \tensor{\Gamma}{^r_r_t} \right) \pd^t + \left( \pd_r \bar{h} + \frac{2}{r} - \tensor{\Gamma}{^r_r_r} \right) \pd^r \right] \eF^\prime 
	. \label{eq:ErrF}
\end{align}
To determine whether the constraints of Tab.~\ref{tab:MTGcc} are satisfied in $\mathfrak{f}(R)$ theories, we substitute $\eF(\bar{R}) = \beta \bar{R}^q$, $\eF^\prime = \beta q \bar{R}^{q-1}$ into Eqs.~\eqref{eq:EttF}--\eqref{eq:ErrF} to obtain
\begin{align}
	\frac{\tensor{\bar{\eE}}{_t_t}}{\beta \lambda} &= 
	q \bar{R}^{q-1} \tensor{\bar{R}}{_t_t} - \frac{1}{2} \bar{R}^q \tensor{\bar{\sg}}{_t_t} + q \left[ \tensor{\bar{\sg}}{_t_t} \left( \pd_t \pd^t + \pd_r \pd^r + \left( \pd_t \bar{h} \right) \pd^t + \left( \pd_r \bar{h} + \frac{2}{r} \right) \pd^r \right) - \pd_t \pd_t + \tensor{\Gamma}{^t_t_t}  \pd_t + \tensor{\Gamma}{^r_t_t} \pd_r \right] \bar{R}^{q-1}
	, \label{eq:EttRq} \\
	\frac{\tensor{\bar{\eE}}{_t^r}}{\beta \lambda} &=
	q \bar{R}^{q-1} \tensor{\bar{R}}{_t^r} - q \left( \pd_t \pd^r + \tensor{\Gamma}{^r_t_t} \pd^t + \tensor{\Gamma}{^r_t_r} \pd^r \right) \bar{R}^{q-1}
	, \label{eq:EtrRq} \\
	\frac{\tensor{\bar{\eE}}{^r^r}}{\beta \lambda} &=
	q \bar{R}^{q-1} \tensor{\bar{R}}{^r^r} - \frac{1}{2} \bar{R}^q \tensor{\bar{\sg}}{^r^r} + q \tensor{\bar{\sg}}{^r^r} \left[ \pd_t \pd^t + \left( \pd_t \bar{h} - \tensor{\Gamma}{^r_r_t} \right) \pd^t + \left( \pd_r \bar{h} + \frac{2}{r} - \tensor{\Gamma}{^r_r_r} \right) \pd^r \right] \bar{R}^{q-1} 
	. \label{eq:ErrRq}
\end{align}
\end{widetext}
Direct evaluation of Eqs.~\eqref{eq:EttRq}--\eqref{eq:ErrRq} using the metric functions Eqs.~\eqref{teq:k0-C}--\eqref{teq:k0-h} and Eqs.~\eqref{teq:k1-C}--\eqref{teq:k1-h} of the $k=0$ and $k=1$ solutions, respectively, shows that the MTG terms $\tensor{\bar{\eE}}{_\mu_\nu}$ of any arbitrary $\mathfrak{f}(R)$ theory conform to the series expansion structures prescribed by Eqs.~\eqref{teq:k0-Ett}--\eqref{teq:k0-Err} and Eqs.~\eqref{teq:k1-Ett}--\eqref{teq:k1-Err} listed in Tab.~\ref{tab:MTGcc}, respectively. In addition, the identities Eqs.~\eqref{teq:k0-cc1cc2}--\eqref{teq:k0-cc3} and Eqs.~\eqref{teq:k1-cc1}--\eqref{teq:k1-cc2} are satisfied identically, i.e.\ no additional conditions are required to be compatible with the semiclassical PBH solutions.

For $k=0$ PBH solutions, the lowest-order MTG coefficients are given explicitly by
\begin{align}
	\ae_{\bar{1}} &= - q c_{12}R_0^{q-3} \sqrt{\bar{\xi}} \Big[ c_{12} \sqrt{\bar{\xi}} \big[ 2 R_0^2 + (q-1) r_\sg \big( 2 R_0 R_1 \nonumber \\
	& \hspace*{-2.7mm} + R_{12} \left( (q-2) R_{12} - R_0 h_{12} \right) \hspace*{-0.5mm} \big) \big] + \left( q-1 \right) r_\sg^2 R_0 R_0^\prime \vphantom{\sqrt{\xi}} \Big]  / \left( 4 r_\sg^3 \right) \nonumber \\
	&= \sqrt{\bar{\xi}} \oe_{\overbar{12}} = \bar{\xi} \o_0 ,
\end{align}
where $R_j$ denote coefficients of the Ricci scalar expansion Eq.~\eqref{teq:k0-R}. Explicit expressions for the MTG coefficients at the next-highest order [i.e.\ those needed to evaluate the third constraint Eq.~\eqref{teq:k0-cc3}] are considerably more convoluted and can be accessed via the {\href{https://github.com/s-murk/MTGcoefficients}{linked Github repository}} \cite{github:MTGcoefficients}.

For the extreme-valued $k=1$ solution, the relevant MTG coefficients of the lowest orders in $x$ are given explicitly by
\begin{align}
	\ae_{\overbar{32}} &= 2^{q-1} c_{32} r_\sg^{-1-2q} \bar{\xi}^3 , \\
	\begin{split}
		\ae_{\bar{1}} &= 2^{q-3} c_{32} r_\sg^{-1-2q} \bar{\xi}^3 \\
		& \qquad \times \left[ 4 h_{12} - 3 q c_{32} \left( 2 q + \left( q - 1 \right) r_\sg^3 R_1 \right) \right] , 
	\end{split}
	\\
	\oe_0 &= 0 , 
	\label{eq:f(R)k1oe0} \\
	\begin{split}
		\oe_{12} &= -3 \times 2^{q-3} q c_{32}^2 r_\sg^{-1-2q} \bar{\xi}^{3/2} \\
		& \hspace*{28mm} \times \left[ 2 q + \left( q - 1 \right) r_\sg^3 R_1 \right] , 
	\end{split}
	\\ 
	\o_{32} &= -2^{q-1} c_{32} r_\sg^{-1-2q} , \\
	\begin{split}
		\o_2 &= 2^{q-3} c_{32} r_\sg^{-1-2q} \\
		& \qquad \times \left[ 4 h_{12} - 3 q c_{32} \left( 2 q + \left( q - 1 \right) r_\sg^3 R_1 \right) \right] ,
	\end{split}
\end{align}
where $R_1$ denotes the coefficient of the $\mathcal{O}(x)$ term in the Ricci scalar expansion Eq.~\eqref{teq:k1-R}. To reduce clutter, we omit the bar label for purely semiclassical functions in what follows.

\section{Physical black holes in fourth-order gravity} \label{sec:fog}

\subsection{Field equations and their constituents}
While $\mathfrak{f}(R)$ gravity theories offer considerably more freedom than GR, they are not the most general MTG with up to fourth-order derivatives in the metric. If one considers up to fourth-order derivatives in the metric, the only possible curvature scalars (in addition to the Ricci scalar $R$) in the gravitational action are the Kretschmann scalar $\mathcal{K} \defeq \tensor{R}{_\mu_\nu_\rho_\sigma} \tensor{R}{^\mu^\nu^\rho^\sigma}$, the square $\tensor{R}{_\mu_\nu} \tensor{R}{^\mu^\nu}$ of the Ricci tensor, and the square $R^2 = \tensor{R}{_\mu^\mu} \tensor{R}{_\nu^\nu}$ of the Ricci scalar\footnote{As pointed out in Ref.~\cite{l:38}, it is technically 
possible to construct two additional invariants that arise from contractions of the Riemann tensor with the completely antisymmetric determinant tensor $\tensor{\delta}{^\mu^\nu^\rho^\sigma} \defeq \tensor{\epsilon}{^\mu^\nu^\rho^\sigma} / \sqrt{\sg}$, where $\tensor{\epsilon}{^\mu^\nu^\rho^\sigma}$ denotes the Levi--Civita symbol in four dimensions. They shall not concern us here as they ultimately do not contribute to the field equations.}. However, it has been demonstrated by virtue of the generalized Chern--Gau{\ss}--Bonnet theorem in four dimensions that only two of these scalars are independent \cite{l:38,s:77,s:78}, resulting in a two-parameter family of field equations. In other words, the quantity
\begin{align}
	\sqrt{-\sg} \mathcal{L}_2 = \sqrt{-\sg} \left( R^2 - 4 \tensor{R}{_\mu_\nu} \tensor{R}{^\mu^\nu} + \tensor{R}{_\mu_\nu_\rho_\sigma} \tensor{R}{^\mu^\nu^\rho^\sigma} \right) 
\end{align}
is a topological invariant and thus vanishes when integrated over spacetimes topologically equivalent to flat space, 
where
\begin{align}
	\mathcal{L}_j \defeq \frac{1}{2^j} \delta^{\mu_1 \cdots \mu_{2j}}_{\nu_1 \cdots \nu_{2j}} R^{\nu_1 \nu_2}_{\mu_1 \mu_2} \cdots R^{\nu_{2j-1} \nu_{2j}}_{\mu_{2j-1} \mu_{2j}}
\end{align}
labels the dimensionally extended Euler densities \cite{bclr:16}. 

Since constant terms in the Lagrangian density do not contribute to the equations of motion, this allows us to eliminate one of the higher-derivative curvature scalars, e.g.\ the Kretschmann scalar $\mathcal{K}$. The gravitational action is then given by
\begin{align}
	\mathcal{S} = \int \sqrt{-\sg} \; \left( - \alpha \tensor{R}{_\mu_\nu} \tensor{R}{^\mu^\nu} + \beta R^2 + \gamma \kappa^{-2} R \right) \; d^4 x ,
\end{align}
where $\kappa^2 = 32 \pi$. It is worth pointing out that theories of the form $\mathfrak{f}(R,\tensor{R}{_\mu_\nu}\tensor{R}{^\mu^\nu},\tensor{R}{_\mu_\nu_\rho_\sigma}\tensor{R}{^\mu^\nu^\rho^\sigma})$ are typically plagued by so-called ghosts: massive states of negative norm that result in an apparent lack of unitarity \cite{sf:10,s:78}. However, models with only $\mathfrak{f}(R,R^2-4\tensor{R}{_\mu_\nu}\tensor{R}{^\mu^\nu}+\tensor{R}{_\mu_\nu_\rho_\sigma}\tensor{R}{^\mu^\nu^\rho^\sigma})$ terms in the action are ghost-free \cite{c:05,nva:06}.

The equations of motion can be derived using the procedure outlined in Ref.~\cite{p:11} and are given by \footnote{They are also provided in Ref.~\cite{s:78} and derived explicitly in Ref.~\cite{w:13}, although some care and diligence is required when comparing expressions from Refs.~\cite{s:78,w:13} with Ref.~\cite{p:11} and those derived here as the former use a different sign convention in the definition of the action and EMT.}
\begin{widetext}
\begin{align}
	\begin{aligned}
	& \alpha \left( - \frac{1}{2} \tensor{R}{_\rho_\sigma} \tensor{R}{^\rho^\sigma} \tensor{\sg}{_\mu_\nu} - \nabla_\nu \nabla_\mu R - 2 \tensor{R}{_\rho_\nu_\mu_\sigma} \tensor{R}{^\sigma^\rho} + \frac{1}{2} \tensor{\sg}{_\mu_\nu} \square R + \square \tensor{R}{_\mu_\nu} \right) \\
	+ \; & \beta \left( 2 R \tensor{R}{_\mu_\nu} - \frac{1}{2} R^2 \tensor{\sg}{_\mu_\nu} - 2 \nabla_\nu \nabla_\mu R + 2 \tensor{\sg}{_\mu_\nu} \square R \right) + \underbrace{\gamma \kappa^{-2} \left( \tensor{R}{_\mu_\nu} - \frac{1}{2} R \tensor{\sg}{_\mu_\nu} \right)}_{\text{GR term}}  = 0 .
	\end{aligned}
	\label{eq:4thOG-EOM}
\end{align}
\end{widetext}
These field equations encode eight dynamical degrees of freedom \cite{s:77,s:78}: two of them correspond to the familiar massless spin-2 graviton, five others to a massive spin-2 particle, and the remaining one to a massive scalar (i.e.\ spin-0) particle. Note that $\beta$ labels contributions of $\mathfrak{f}(R)$ gravity [cf.\ Sec.~\ref{sec:f(R)}, Eq.~\eqref{eq:E_mu_nu}] for the parameter choice $q=2$, i.e.\ $\eF(R) = \beta R^2$, $\eF^\prime = 2 \beta R$, which corresponds to the Starobinsky model \cite{aaS:79,aaS:80}. We make no assumptions about the parameters $\alpha$, $\beta$, and $\gamma$. In practice, their values must be tuned to satisfy various experimental constraints. The equations of motion can be conveniently recast in the form
\begin{align}
	\tensor{\eE}{_\mu_\nu} + \gamma \kappa^{-2} \left( \tensor{R}{_\mu_\nu} - \frac{1}{2} R \tensor{\sg}{_\mu_\nu} \right) = 0 ,
\end{align}
with the MTG terms $\tensor{\eE}{_\mu_\nu}$ (i.e.\ the deviations from the semiclassical Einstein equations) given by
\begin{widetext}
\begin{align}
	\begin{aligned}
		\tensor{\eE}{_\mu_\nu} &= - \left( \alpha + 2 \beta \right) \nabla_\nu \nabla_\mu R + \left( \frac{\alpha}{2} + 2 \beta \right) \tensor{\sg}{_\mu_\nu} \square R \\
		& \qquad + \alpha \left( - \frac{1}{2} \tensor{R}{_\rho_\sigma} \tensor{R}{^\rho^\sigma} \tensor{\sg}{_\mu_\nu} - 2 \tensor{R}{_\rho_\nu_\mu_\sigma} \tensor{R}{^\rho^\sigma} + \square \tensor{R}{_\mu_\nu} \right) + \beta \left( 2 R \tensor{R}{_\mu_\nu} - \frac{1}{2} R^2 \tensor{\sg}{_\mu_\nu} \right) .
	\end{aligned}
	\label{eq:gen_4thOG_MTG_term}
\end{align}
All terms in Eq.~\eqref{eq:gen_4thOG_MTG_term} contain fourth-order derivatives. In spherical symmetry, the relevant MTG terms are
\begin{align}	
	\begin{split}
		\tensor{\eE}{_t_t} &= - \left( \alpha + 2 \beta \right) \nabla_t \nabla_t R + \left( \frac{\alpha}{2} + 2 \beta \right) \tensor{\sg}{_t_t} \square R \\
		& \qquad + \alpha \left( - \frac{1}{2} \tensor{R}{_\rho_\sigma} \tensor{R}{^\rho^\sigma} \tensor{\sg}{_t_t} - 2 \tensor{R}{_\rho_t_t_\sigma} \tensor{R}{^\rho^\sigma} + \square \tensor{R}{_t_t} \right) + \beta \left( 2 R \tensor{R}{_t_t} - \frac{1}{2} R^2 \tensor{\sg}{_t_t} \right) ,
	\end{split}
	\label{eq:4thOG_Ett}
	\\
	\tensor{\eE}{_t^r} &= - \left( \alpha + 2 \beta \right) \nabla_t \nabla^r R + \alpha \left( - 2 \tensor{R}{_\rho_t^r_\sigma} \tensor{R}{^\rho^\sigma} + \square \tensor{R}{_t^r} \right) + 2 \beta R \tensor{R}{_t^r} , 
	\label{eq:4thOG_Etr} \\
	\begin{split}
		\tensor{\eE}{^r^r} &= - \left( \alpha + 2 \beta \right) \nabla^r \nabla^r R + \left( \frac{\alpha}{2} + 2 \beta \right) \tensor{\sg}{^r^r} \square R \\
		& \qquad + \alpha \left( - \frac{1}{2} \tensor{R}{_\rho_\sigma} \tensor{R}{^\rho^\sigma} \tensor{\sg}{^r^r} - 2 \tensor{R}{_\rho^r^r_\sigma} \tensor{R}{^\rho^\sigma} + \square \tensor{R}{^r^r} \right) + \beta \left( 2 R \tensor{R}{^r^r} - \frac{1}{2} R^2 \tensor{\sg}{^r^r} \right) .
	\end{split}
	\label{eq:4thOG_Err} 
\end{align}
\end{widetext}
Note that here, unlike $\beta$ in Sec.~\ref{sec:f(R)} [cf.\ Eqs.~\eqref{eq:EttF}--\eqref{eq:ErrRq}], the coefficients $\alpha$ and $\beta$ are not absorbed into the lhs of the equations for the MTG terms $\tensor{\eE}{_\mu_\nu}$, which allows us to easily distinguish their respective contributions. The Ricci and Riemann tensor contractions that appear in Eqs.~\eqref{eq:4thOG_Ett} and \eqref{eq:4thOG_Err} decompose into
\begin{align}
	\begin{split}
		\tensor{R}{_\rho_\sigma} \tensor{R}{^\rho^\sigma} &= \tensor{R}{_t_t} \tensor{R}{^t^t} + 2 \tensor{R}{_t_r} \tensor{R}{^t^r} + \tensor{R}{_r_r} \tensor{R}{^r^r} \\
		& \qquad + \tensor{R}{_\theta_\theta} \tensor{R}{^\theta^\theta} + \tensor{R}{_\phi_\phi} \tensor{R}{^\phi^\phi} , 
	\end{split}
	\\
	\tensor{R}{_\rho_t_t_\sigma} \tensor{R}{^\rho^\sigma} &= \tensor{R}{_r_t_t_r} \tensor{R}{^r^r} + \tensor{R}{_\theta_t_t_\theta} \tensor{R}{^\theta^\theta} + \tensor{R}{_\phi_t_t_\phi} \tensor{R}{^\phi^\phi} , \\
	\tensor{R}{_\rho^r^r_\sigma} \tensor{R}{^\rho^\sigma} &= \tensor{R}{_t^r^r_t} \tensor{R}{^t^t} + \tensor{R}{_\theta^r^r_\theta} \tensor{R}{^\theta^\theta} + \tensor{R}{_\phi^r^r_\phi} \tensor{R}{^\phi^\phi} .
\end{align}
The second-order covariant derivatives of the Ricci scalar and its d'Alembertian can be computed straightforwardly using Eqs.~\eqref{eq:sec_cov_der} and \eqref{eq:d'Alembertian}, respectively. To evaluate d'Alembertians of the Ricci tensor, we first note that the covariant derivatives of a $\left(0,2\right)$, $\left(1,1\right)$, and $\left(2,0\right)$ tensor field $\tensor{R}{_\mu_\nu}$, $\tensor{R}{_\mu^\nu}$, and $\tensor{R}{^\mu^\nu}$ with respect to $\alpha$ are given by
\begin{align}
	\nabla_\alpha \tensor{R}{_\mu_\nu} &= \partial_\alpha \tensor{R}{_\mu_\nu} - \tensor{\Gamma}{^\zeta_\alpha_\mu} \tensor{R}{_\zeta_\nu} - \tensor{\Gamma}{^\zeta_\alpha_\nu} \tensor{R}{_\mu_\zeta} , \label{eq:covD(0,2)} \\
		\nabla_\alpha \tensor{R}{_\mu^\nu} &= \partial_\alpha \tensor{R}{_\mu^\nu} - \tensor{\Gamma}{^\zeta_\alpha_\mu} \tensor{R}{_\zeta^\nu} + \tensor{\Gamma}{^\nu_\alpha_\zeta} \tensor{R}{_\mu^\zeta} , \label{eq:covD(1,1)} \\
	\nabla_\alpha \tensor{R}{^\mu^\nu} &= \partial_\alpha \tensor{R}{^\mu^\nu} + \tensor{\Gamma}{^\mu_\alpha_\zeta} \tensor{R}{^\zeta^\nu} + \tensor{\Gamma}{^\nu_\alpha_\zeta} \tensor{R}{^\mu^\zeta} , \label{eq:covD(2,0)}
\end{align}
where $\tensor{\Gamma}{^\zeta_\mu_\nu}$ denotes Christoffel symbols of the second kind. Thus the relevant second-order covariant derivatives of the Ricci tensor take the form
\begin{widetext}
\begin{align}
	\begin{aligned}
		\nabla_\alpha \nabla_\beta \tensor{R}{_\mu_\nu} &= \nabla_\alpha \left( \pd_\beta \tensor{R}{_\mu_\nu} - \tensor{\Gamma}{^\zeta_\mu_\beta} \tensor{R}{_\zeta_\nu} - \tensor{\Gamma}{^\zeta_\beta_\nu} \tensor{R}{_\mu_\zeta} \right) \\
		&= \pd_\alpha \left( \pd_\beta \tensor{R}{_\mu_\nu} - \tensor{\Gamma}{^\zeta_\mu_\beta} \tensor{R}{_\zeta_\nu} - \tensor{\Gamma}{^\zeta_\beta_\nu} \tensor{R}{_\mu_\zeta} \right) - \tensor{\Gamma}{^\chi_\beta_\alpha} \left( \pd_\chi \tensor{R}{_\mu_\nu} - \tensor{\Gamma}{^\zeta_\mu_\chi} \tensor{R}{_\zeta_\nu} - \tensor{\Gamma}{^\zeta_\chi_\nu} \tensor{R}{_\mu_\zeta} \right) \\
		& \qquad - \tensor{\Gamma}{^\chi_\mu_\alpha} \left( \pd_\beta \tensor{R}{_\chi_\nu} - \tensor{\Gamma}{^\zeta_\chi_\beta} \tensor{R}{_\zeta_\nu} - \tensor{\Gamma}{^\zeta_\beta_\nu} \tensor{R}{_\chi_\zeta} \right) - \tensor{\Gamma}{^\chi_\nu_\alpha} \left( \pd_\beta \tensor{R}{_\mu_\chi} - \tensor{\Gamma}{^\zeta_\mu_\beta} \tensor{R}{_\zeta_\chi} - \tensor{\Gamma}{^\zeta_\beta_\chi} \tensor{R}{_\mu_\zeta} \right) ,
	\end{aligned}
\end{align}
\end{widetext}
and analogously for the mixed $(1,1)$ and contravariant $(2,0)$ Ricci tensor components. In spherical symmetry, $\tensor{\sg}{^\mu^\nu} = 0$ for $\mu \neq \nu$ due to the diagonal form of the metric tensor [cf.\ Eq.~\eqref{eq:metric}], and thus the d'Alembertian $\square \defeq \tensor{\sg}{^\mu^\nu} \nabla_\mu \nabla_\nu$ simplifies to $\square \equiv \tensor{\sg}{^\mu^\mu} \nabla_\mu \nabla_\mu$. In addition, the Ricci tensor and Christoffel symbols of the second kind are symmetric in their covariant indices. Therefore, all covariant derivatives with respect to the angular variables $\lbrace \theta , \phi \rbrace$ vanish:
\begin{align}
	\begin{split}
		& \pd_\theta \tensor{R}{_t_t} = \pd_\phi \tensor{R}{_t_t} = 0 \; , \; \tensor{R}{_\chi_t} = 0 \; \; \text{for} \; \; \chi \in \lbrace \theta , \phi \rbrace \\
		& \quad \text{and} \; \; \tensor{\Gamma}{^t_\zeta_t} = \tensor{\Gamma}{^r_\zeta_t} = 0 \; \; \text{for} \; \; \zeta \in \lbrace \theta , \phi \rbrace \\
		& \qquad  \Rightarrow \; \; \nabla_\theta \tensor{R}{_t_t} = \nabla_\phi \tensor{R}{_t_t} = 0 \; , 
	\end{split}		
	\\[4.1mm]
	\begin{split}
		& \pd_\theta \tensor{R}{_t^r} = \pd_\phi \tensor{R}{_t^r} = 0 \; , \; \tensor{R}{_\chi^r} = \tensor{R}{_t^\chi} = 0 \; \; \text{for} \; \; \chi \in \lbrace \theta , \phi \rbrace \\
		& \quad \text{and} \; \; \tensor{\Gamma}{^t_\zeta_t} = \tensor{\Gamma}{^r_\zeta_t} = \tensor{\Gamma}{^r_\zeta_r} = 0 \; \; \text{for} \; \; \zeta \in \lbrace \theta , \phi \rbrace \\
		& \qquad \Rightarrow \; \; \nabla_\theta \tensor{R}{_t^r} = \nabla_\phi \tensor{R}{_t^r} = 0 \; ,
	\end{split}
	\\
	\begin{split}
		& \pd_\theta \tensor{R}{^r^r} = \pd_\phi \tensor{R}{^r^r} = 0 \; , \; \tensor{R}{^\chi^r} = 0 \; \; \text{for} \; \; \chi \in \lbrace \theta , \phi \rbrace \\
		& \quad \text{and} \; \; \tensor{\Gamma}{^r_\zeta_t} = \tensor{\Gamma}{^r_\zeta_r} = 0 \; \; \text{for} \; \; \zeta \in \lbrace \theta , \phi \rbrace \\ 
		& \qquad \Rightarrow \; \; \nabla_\theta \tensor{R}{^r^r} = \nabla_\phi \tensor{R}{^r^r } = 0 \; .
	\end{split}
\end{align}
The d'Alembertians of the relevant Ricci tensor components are therefore given by
\begin{align}
	\square \tensor{R}{_t_t} &= \tensor{\sg}{^t^t} \nabla_t \nabla_t \tensor{R}{_t_t} + \tensor{\sg}{^r^r} \nabla_r \nabla_r \tensor{R}{_t_t} , 
	\label{eq:dAlRtt} \\
	\square \tensor{R}{_t^r} &= \tensor{\sg}{^t^t} \nabla_t \nabla_t \tensor{R}{_t^r} + \tensor{\sg}{^r^r} \nabla_r \nabla_r \tensor{R}{_t^r} , 
	\label{eq:dAlRtr} \\
	\square \tensor{R}{^r^r} &= \tensor{\sg}{^t^t} \nabla_t \nabla_t \tensor{R}{^r^r} + \tensor{\sg}{^r^r} \nabla_r \nabla_r \tensor{R}{^r^r} .
	\label{eq:dAlRrr}
\end{align}
To determine whether the generic fourth-order gravity theory specified by the field equations Eq.~\eqref{eq:4thOG-EOM} is compatible with the PBHs of semiclassical gravity, we compute its MTG terms Eqs.~\eqref{eq:4thOG_Ett}--\eqref{eq:4thOG_Err} and check if the constraints summarized in Tab.~\ref{tab:MTGcc} are satisfied.

\subsection{$k=0$ PBH solutions}
First, we evaluate Eqs.~\eqref{eq:4thOG_Ett}--\eqref{eq:4thOG_Err} using the $k=0$ metric functions Eqs.~\eqref{teq:k0-C}--\eqref{teq:k0-h}. The resulting lowest-order coefficients are given in Eq.~\eqref{eq:MTGc_k0_lowestO_4thOG}. Again, the explicit expressions for coefficients at the next-highest order are considerably more convoluted and are therefore provided in the {\href{https://github.com/s-murk/MTGcoefficients}{linked Github repository}} \cite{github:MTGcoefficients}. All three MTG terms conform to the series expansion structures prescribed by Eqs.~\eqref{teq:k0-Ett}--\eqref{teq:k0-Err}, and the identities Eqs.~\eqref{teq:k0-cc1cc2}--\eqref{teq:k0-cc3} between their coefficients are satisfied identically, i.e.\ without any additional constraints. Consequently, generic fourth-order gravity theories are compatible with the $k=0$ PBH solutions of semiclassical gravity, which can be regarded as zeroth-order terms in pertubative solutions of such theories.

It is worth noting that the constraints are satisfied individually by the $\alpha$ and $\beta$ terms, i.e.\ they are satisfied in both limits $\alpha \to 0$ and $\beta \to 0$. This behavior is expected since the $\beta$ terms describe a specific type of $\mathfrak{f}(R)$ theory, and we have demonstrated in Sec.~\ref{sec:f(R)} that all $\mathfrak{f}(R)$ theories generically satisfy all constraints. Consequently, if the constraints are satisfied individually by the $\beta$ terms (i.e.\ in the limit $\alpha \to 0$), they must also be satisfied individually by the $\alpha$ terms (i.e.\ in the limit $\beta \to 0$) in order for the entire fourth-order theory to satisfy all constraints.

\subsection{$k=1$ PBH solution}
Evaluation of Eqs.~\eqref{eq:4thOG_Ett}--\eqref{eq:4thOG_Err} with the metric functions Eqs.~\eqref{teq:k1-C}--\eqref{teq:k1-h} of the extreme-valued $k=1$ solution yields the following coefficients at the two lowest orders:
\begin{align}
	\ae_{\overbar{32}} &= \left( \alpha + 2 \beta \right) c_{32} \xi^3 / r_\sg^5 , 
	\label{eq:4thOG_k1_AE32} \\
	\ae_{\bar{1}} &= \left( \alpha + 2 \beta \right) c_{32} \xi^3 \left( 2 h_{12} - 3 c_{32} \left( 4 + r_\sg^3 R_1 \right) \right) / \left( 2 r_\sg^5 \right) , 
	\label{eq:4thOG_k1_AE1} \\
	\oe_0 &= 0 , 
	\label{eq:4thOG_k1_oe0} \\
	\oe_{12} &= - \left( \alpha + 2 \beta \right) c_{32}^2 \xi^{3/2} \left( 4 + r_\sg^3 R_1 \right) / \left( 2 r_\sg^5 \right) , 
	\label{eq:4thOG_k1_oe12} \\
	\o_{32} &= - \left( \alpha + 2 \beta \right) c_{32} / r_\sg^5 , 
	\label{eq:4thOG_k1_o32} \\
	\o_2 &= \left( \alpha + 2 \beta \right) c_{32} \left( 2 h_{12} - 3 c_{32} \left( 4 + r_\sg^3 R_1 \right) \right) / \left( 2 r_\sg^5 \right) .
	\label{eq:4thOG_k1_o2}
\end{align}
Once again, the MTG terms match the required expansion structures of Eqs.~\eqref{teq:k1-Ett}--\eqref{teq:k1-Err}, and the two relations Eqs.~\eqref{teq:k1-cc1}--\eqref{teq:k1-cc2} between their coefficients are satisfied identically. We note that, even in the more general fourth-order theory, the coefficient $\oe_0 = 0$ as in $\mathfrak{f}(R)$ gravity [cf.\ Eq.~\eqref{eq:f(R)k1oe0} in Sec.~\ref{sec:f(R)}], even though a term of order $\mathcal{O}(x^0)$ in $\tensor{\eE}{_t^r}$ would not tarnish the compatibility with the semiclassical $k=1$ PBH solution according to Eq.~\eqref{teq:k1-Etr}.

From Eqs.~\eqref{eq:4thOG_k1_AE32}--\eqref{eq:4thOG_k1_o2}, we see that the $k=1$ constraints are not only satisfied individually (as in the $k=0$ case) by the $\alpha$ and $\beta$ terms that label modifications of the semiclassical equations of motion, but in fact their contributions to the MTG coefficients are exactly the same. This is not true for MTG terms arising from the $k=0$ metric, where the $\alpha$ and $\beta$ terms of Eq.~\eqref{eq:4thOG-EOM} lead to distinct contributions [see, for instance, Eq.~\eqref{eq:MTGc_k0_lowestO_4thOG}, which includes pure $\alpha$ and pure $\beta$ contributions in addition to mixed $(\alpha+2\beta)$ terms], but it is true generically for the $k=1$ metric, i.e.\ for all MTG terms $\tensor{\eE}{_\mu_\nu}$ and all of their coefficients, even those at higher orders. This implies that the extreme-valued $k=1$ solution that describes the formation of PBHs gives rise to the same MTG terms in both theories and should be interpreted as an indication that the process of PBH formation in fourth-order MTG may not differ from that of semiclassical gravity, despite the availability of additional gravitational degrees of freedom. Contributions of the modifications labeled by $\alpha$ and $\beta$ to the energy density $E$ and pressure $P$ are also the same. Close to the horizon, their leading contributions scale as $\sim 1/r_\sg^4$, whereas the leading GR contribution scales as $\sim 1/r_\sg^2$.

\section{Discussion} \label{sec:disc}
MTG must satisfy several constraints (summarized in Tab.~\ref{tab:MTGcc}) to be compatible with the two classes of PBHs in semiclassical gravity (whose properties are summarized in Tab.~\ref{tab:PBHsol}). In $\mathfrak{f}(R)$ and generic fourth-order gravity theories, all of the constraints are satisfied identically, and the semiclassical PBH solutions can be regarded as zeroth-order terms in perturbative solutions of these models. Consequently, the formation of an apparent horizon on its own is not sufficient to distinguish between semiclassical gravity and families of fourth-order MTG. Only a detailed analysis of the response of the near-horizon geometry to perturbations may allow us to identify potential observationally distinguishable features.

In addition to the fact that their MTG terms follow the expansion structures of Eqs.~\eqref{teq:k0-Ett}--\eqref{teq:k0-Err} and Eqs.~\eqref{teq:k1-Ett}--\eqref{teq:k1-Err} and satisfy the coefficient relations Eqs.~\eqref{teq:k0-cc1cc2}--\eqref{teq:k0-cc3} and Eqs.~\eqref{teq:k1-cc1}--\eqref{teq:k1-cc2} for $k=0$ and $k=1$ PBH solutions, respectively, a generic feature of fourth-order gravity theories is that $\oe_0 = 0$ for the extreme-valued $k=1$ solution, i.e.\ there is no $\mathcal{O}(x^0)$ term in the MTG term $\tensor{\eE}{_t^r}$ of the $tr$ equation [cf.\ Eq.~\eqref{teq:k1-Etr} in Tab.~\ref{tab:MTGcc}]. Moreover, the MTG terms arising from the $k=1$ metric that describes PBH formation in semiclassical gravity appear to be independent of the additional degrees of freedom provided by the $\alpha$ terms in the equations of motion Eq.~\eqref{eq:4thOG-EOM} of generic fourth-order gravity theories.

In future works, we plan to extend the analysis of Ref.~\cite{mt:21a} by systematically investigating PBH formation scenarios in more general settings, e.g.\ by considering non-spherically symmetric spacetimes and including angular momentum, as well as to extend the analysis presented here by investigating the constraints in additional modified gravity theories, such as the recent reformulation of Gau{\ss}--Bonnet gravity in four dimensions with non-trivial gravitational dynamics \cite{gl:20,hkmp:20}.

\acknowledgments
I would like to thank Robert Mann for drawing my attention to reformulations of Gau{\ss}--Bonnet gravity and Daniel Terno for useful discussions and helpful comments. This work was supported by an International Macquarie University Research Excellence Scholarship and a Sydney Quantum Academy Scholarship.

\onecolumngrid
\appendix

\section{MTG term coefficients of $k=0$ PBH solutions in fourth-order gravity theories} \label{app:MTG_coefficients_4thOG}
Evaluation of Eqs.~\eqref{eq:4thOG_Ett}--\eqref{eq:4thOG_Err} with the $k=0$ metric functions Eqs.~\eqref{teq:k0-C}--\eqref{teq:k0-h} yields the following explicit expressions for the lowest-order MTG coefficients of Eqs.~\eqref{teq:k0-Ett}--\eqref{teq:k0-Err} in fourth-order gravity theories: 
\begin{align}
	\begin{aligned}
		\ae_{\bar{1}} &= \frac{1}{24 r_\sg^5} \bigg[ - 6 \alpha (c_1-1)^3 r_\sg \xi + 2 \alpha c_{12}^3 \xi \Big( \hspace*{-1mm} -12 h_{12} + \left( 15 h_{12} h_1 + h_{12}^3 - 21 h_{32} \right) r_\sg \Big) \\
		& \qquad + 6 c_{12}^2 \xi \bigg( 5 \alpha + \alpha c_1 \left( 2 r_\sg \left( h_1 - h_{12}^2 \right) - 9 \right) + r_\sg \Big( \alpha \left( 4 c_2 - 2 h_1 - c_{32} h_{12} + 2 h_{12}^2 \right) - 4 \beta r_\sg R_0 \\
		& \qquad - 2 \left( \alpha + 2 \beta \right) r_\sg^2 R_1 + \left( \alpha + 2 \beta \right) h_{12} r_\sg^2 R_{12} \Big) \bigg) + 12 \alpha (c_1-1) r_\sg^2 \sqrt{\xi} c_{12}^\prime - 12 \alpha c_{12}^2 r_\sg^2 \sqrt{\xi} h_{12}^\prime \\
		& \qquad + 3 c_{12} r_\sg \sqrt{\xi} \left( \alpha \sqrt{\xi} (c_1-1) \Big( h_{12} (c_1-1) - 4 c_{32} \Big) - 4 \alpha h_{12} r_\sg h_{12}^\prime - 2 \left( \alpha + 2 \beta \right) r_\sg^3 R_0^\prime \right) \bigg] \\
		&= \sqrt{\xi} \oe_{\overbar{12}} = \xi \o_0 .
	\end{aligned}
	\label{eq:MTGc_k0_lowestO_4thOG}
\end{align}
Explicit expressions for the next-highest order coefficients $\ae_{\overbar{12}}$, $\oe_0$, and $\o_{12}$ of Eqs.~\eqref{teq:k0-Ett}--\eqref{teq:k0-Err} are provided in the {\href{https://github.com/s-murk/MTGcoefficients}{linked Github repository}} \cite{github:MTGcoefficients}.

\twocolumngrid

\end{document}